\documentclass[cits]{PoS-pdf}

\providecommand{\href}[2]{#2}
\addcontentsline{toc}{section}{References}
\newcommand{\secdec}{{\textsc{SecDec}}}

\newcommand{\bea}{\begin{eqnarray}}
\newcommand{\eea}{\end{eqnarray}\noindent}

\def\eps{\epsilon}
\def\be{\begin{align}}
\def\ee{\end{align}}
\def\bea{\begin{align}}
\def\eea{\end{align}}

\title{Numerical evaluation of massive multi-loop integrals with SecDec}

\ShortTitle{Numerical evaluation of massive multi-loop integrals with SecDec}

\author{S.~Borowka, \speaker{G.~Heinrich}\\
 Max-Planck-Institute for Physics,  Munich, Germany\\
 E-mail: \email{\{sborowka,gudrun\}@mpp.mpg.de}}

\abstract{
The program package \secdec{} is presented, allowing the numerical 
evaluation of multi-loop integrals. 
The restriction to Euclidean kinematics of version 1.0 has been lifted: 
thresholds can be handled by an automated deformation of 
the integration contour into the complex plane. 
Other new features of the program, which go beyond the standard decomposition 
of loop integrals, are also described.
The program is publicly available at http://secdec.hepforge.org.}

\FullConference{Loops and Legs in Quantum Field Theory, 
11th DESY Workshop on Elementary Particle Physics \\
                 April 15-20, 2012\\
                 Wernigerode, Germany}

\begin{document}

\section{Introduction}

These days  the amount and the precision of high energy collider physics data is 
increasing rapidly, and the properties of a new discovery 
have to be studied as detailed as possible. Therefore  precise theory predictions 
are of major importance.
While for many observables next-to-leading-order (NLO) precision is 
sufficient, there is a considerable number of examples where 
corrections beyond NLO are required. 
In these proceedings we present the program \secdec{}\cite{Carter:2010hi,Borowka:2012yc}, 
which can assist the numerical calculation 
of such corrections in a process independent way, providing examples for 
multi-loop integrals as well as dimensionally regularised phase space integrals.
The program is based on the method of sector decomposition\,\cite{Binoth:2000ps,Roth:1996pd,Hepp:1966eg},
which is an algorithm to factorise the poles in the regularisation parameter $\eps$ from 
complicated multi-parameter integrals. Other public implementations of sector decomposition can be 
found in \cite{Bogner:2007cr,Smirnov:2008py,Smirnov:2009pb,Gluza:2010rn}. 

While the program \secdec{}-1.0\,\cite{Carter:2010hi} and the public programs mentioned above 
are limited to  kinematics where the values of the Mandelstam invariants and 
masses have to be such that the denominator of the integrand is guaranteed to be of definite sign, 
this restriction is lifted in \secdec{}-2.0\,\cite{Borowka:2012yc}.
Integrable singularities located on the real axis are avoided by an automated deformation of the 
integration contour into the complex plane.
The method of contour deformation in a multi-dimensional parameter space  
has been pioneered in\,\cite{Soper:1999xk} and later has been refined and 
applied to various  calculations at one loop\,\cite{Binoth:2002xh,Binoth:2005ff,Nagy:2006xy,Gong:2008ww,Lazopoulos:2007ix,Becker:2010ng,Becker:2011vg} 
and at two loops\,\cite{Kurihara:2005ja,Anastasiou:2007qb,Anastasiou:2008rm,Freitas:2012iu}.\\
Numerical methods using dispersion relations, numerical extrapolation, differential equations and/or 
numerical integration of Mellin-Barnes representations  also 
have been worked out, see e.g.\,\cite{Bauberger:1994by,Caffo:2002ch,Pozzorini:2005ff,Czakon:2005rk,Czakon:2008zk,Freitas:2010nx,Yuasa:2011ff}. 
However, most of these methods or programs are limited either to specific classes of integrals, or the parameters 
tuning the numerical integration have to be adapted carefully by the authors in an iterative procedure. 

For the program \secdec{}-2, the aim was to offer a package which is ``multi-purpose", 
i.e.  which can handle very different types of multi-loop/multi-scale integrals in a uniform setup.
This universality may come at the expense of sub-optimal performance as compared to dedicated programs
for specific classes of diagrams, but on the other hand offers a tool of very wide applicability.

\section{Main features of the program \secdec{} }

The program performs the following task: it turns a parameter integral, which can contain 
ultraviolet and/or infrared singularities regulated by a parameter $\epsilon$, 
into a Laurent series in $\epsilon$, where the coefficients are calculated numerically.
For multi-loop integrals, integrable singularities (e.g. due to kinematic thresholds) can also be handled.
To start with, the program turns the user information about a Feynman graph 
(i.e. number of legs, loops, propagators, vertices, on-shell conditions) automatically into the corresponding 
Feynman parameter representation.
The extraction of the poles in $1/\epsilon$ is purely algebraic. 
The coefficients of the poles are sets of finite parameter integrals, which are integrated numerically.

\subsection{General form of multi-loop integrals}
Here we will focus on applications of \secdec{}-2 to multi-loop integrals, 
for details about more general parameter integrals we refer to \cite{Carter:2010hi,Borowka:2012yc}.
After Feynman parametrisation, a scalar Feynman integral $G$ in $D$ dimensions 
at $L$ loops with  $N$ propagators, where 
the propagators can have arbitrary, not necessarily integer powers $\nu_j$,  
can be written as
\begin{eqnarray}
G&=&\frac{(-1)^{N_{\nu}}}{\prod_{j=1}^{N}\Gamma(\nu_j)}\Gamma(N_{\nu}-LD/2)\int
\limits_{0}^{\infty} 
\,\prod\limits_{j=1}^{N}dx_j\,\,x_j^{\nu_j-1}\,\delta(1-\sum_{l=1}^N x_l)
\frac{{\cal U}(\vec x)^{N_{\nu}-(L+1) D/2}}
{{\cal F}(\vec x)^{N_\nu-L D/2}}\;,\label{eq:defInt}
\end{eqnarray}
with $N_\nu=\sum_{j=1}^N\nu_j$.
The functions ${\cal U}$ and ${\cal F}$ can be constructed either directly 
from the momentum representation or from the topology of the corresponding 
Feynman graph~\cite{Smirnov:2006ry,Heinrich:2008si}. 
The implementation of the construction based on the topology only, without the need to specify 
the propagators in terms of loop momenta, is one of the new features of  \secdec{}-2. 
${\cal U}$ is a positive semi-definite function, whose  
vanishing  is related to the  UV subdivergences of the graph. 
In the region where all Lorentz invariants formed from external momenta are negative, 
which we will call the {\em Euclidean region}, 
${\cal F}$ is also a positive semi-definite function 
of the Feynman parameters $x_j$ and the invariants.  
If some of the invariants are zero, for example if some of the external momenta
are light-like, the vanishing of  ${\cal F}$  may induce an IR divergence.
Thus it depends on the {\em kinematics}
and not only on the topology (like in the UV case) 
whether a zero of ${\cal F}$ leads to a divergence or not. Therefore general theorems about the 
IR singularity structure of multi-loop integrals are sparse, 
but for practical purposes sector decomposition can provide information about the singularity structure
and numerical results, because it offers a constructive algorithm to extract the poles in $1/\eps$.

\subsection{Deformation of the integration contour}
\label{sec:contour}
As mentioned already, the integrand in eq.~(\ref{eq:defInt}) can diverge for certain values of 
kinematical invariants and Feynman parameters. In cases where this 
corresponds to an integrable singularity of logarithmic or square root type, related 
to normal thresholds, we can make use of Cauchy's theorem to
avoid the  poles on the real axis by a deformation of the integration contour 
into the complex plane. 
As long as the deformation is in accordance with the causal $i\delta$  prescription 
of the Feynman propagators, and no poles are crossed while changing the integration
path, the integration contour can be altered such that
the convergence of the numerical integration 
is assured. 
The $i\delta$ prescription
tells us that the contour deformation should be such that 
the imaginary part of ${\cal F}$ is 
always  negative. 
For real masses and Mandelstam invariants $s_{ij}$, the following ansatz\,\cite{Soper:1999xk,Binoth:2005ff}
is therefore convenient:
\begin{eqnarray}
\label{eq:condef}
\vec{z}( \vec x) = \vec{x} - i\;  \vec{\tau}(\vec{x})\;\; , \;\;
\tau_k = \lambda\, x_k (1-x_k)
 \, \frac{\partial {\cal F}(\vec{x})}{\partial x_k}  \;.
\end{eqnarray}
Unless we are faced with a leading Landau singularity where both ${\cal F}$ and its
derivatives with respect to $x_i$ vanish, 
the deformation leads to a well behaved integral at the points where the function ${\cal F}$ vanishes. 
In terms of the new variables, we thus obtain
\begin{equation}
\label{eq:newF}
{\cal F}(\vec{z}(\vec{x}))={\cal F}(\vec{x})
-i\,\lambda\,\sum\limits_{j} \, x_j (1-x_j)\, \left(\frac{\partial {\cal F}}{\partial x_j}  \right)^2 + {\cal O}(\lambda^2)\;,
\end{equation}
such that ${\cal F}$ acquires a negative imaginary part of order $\lambda$. 
The size of $\lambda$ determines the scale of the deformation. 
The initial value for the parameter $\lambda$ can be given by the user;
however, the program tries to optimise the value for $\lambda$ for each 
subsector function 
and may overwrite the user's choice if the latter is found to be inconvenient.

\subsection{Installation and usage of the program}

The program can be downloaded from 
{\tt http://secdec.hepforge.org}.
Unpacking the tar archive via 
{\it  tar xzvf SecDec.tar.gz} will create a directory called {\tt SecDec}. 
Running  {\it ./install} in the {\tt SecDec} directory
will install the package.
Prerequisites are Mathematica, version 6 or above, Perl 
(installed by default on most Unix/Linux systems), 
a Fortran compiler,  or a C++ 
compiler if the C++ option is used. If contour deformation is required, the C++ option must be used.
The libraries {\small CUBA}\,\cite{Hahn:2004fe,Agrawal:2011tm} and {\small BASES}\,\cite{Kawabata:1995th} 
which are used for the numerical integration come with the package \secdec{}.

\subsection*{Usage}

\begin{enumerate}

\item Change to the subdirectory {\tt loop} or {\tt general}, depending 
on whether you would like to calculate a loop integral or a more general parameter integral.
\item Copy the files {\tt param.input} and {\tt template.m} to 
create your own parameter and template files  {\tt myparamfile.input}, {\tt mytemplatefile.m}.
\item Set the desired parameters in {\tt myparamfile.input} and specify the 
Feynman graph or the function to evaluate in {\tt mytemplatefile.m}.
\item Execute the command {\it ./launch -p myparamfile.input -t mytemplatefile.m} 
in the shell.  \\
If you omit the option {\it -p myparamfile.input}, the file {\tt param.input} will be taken as default.
Likewise, if you omit the option {\it -t mytemplatefile.m}, 
the file {\tt template.m} will be taken as default.
If your files {\tt myparamfile.input, mytemplatefile.m} are in a different directory, say, 
{\it myworkingdir}, 
 use the option {\it -d myworkingdir}, i.e. the full command then looks like 
 {\it ./launch -d myworkingdir -p myparamfile.input -t mytemplatefile.m}, 
 executed from the directory {\tt SecDec/loop} or
 {\tt SecDec/general}. 
\item Collect the results. If the calculations are done sequentially on a single machine, 
    the results will be collected automatically.
If the jobs have been submitted to a cluster,    
	when all jobs have finished, use  the command 
	{\it ./results.pl [-d myworkingdir -p myparamfile]}. 
In both cases, the files containing the final results will be located in the {\tt graph} subdirectory
	specified in the input file.

\end{enumerate}

\section{New features of the program}

Version 2 of  \secdec{} contains the following new features.
\begin{enumerate}
\item The most important new feature is the fact that multi-scale loop integrals can now 
be evaluated without restricting the kinematics to the Euclidean region. 
\item For scalar multi-loop integrals, the integrand can be constructed from the topology of 
the diagram, so the user only has to provide the vertices and the propagator masses, but 
does not have to provide the momentum flow anymore.
\item The files for the numerical integration  of multi-scale loop 
diagrams with contour deformation are written in C++ rather than Fortran. 
\item A parallelisation of the algebraic part for Mathematica versions 7 and higher is possible 
if several cores are available. Parallelisation of the numerical part is also possible.
\item  The so-called {\it primary sector decomposition}\,\cite{Binoth:2000ps} to eliminate 
 the constraint $\delta\left(1-\sum_i x_i \right)$ on the Feynman parameters, see eq.\,(\ref{eq:defInt}), can be skipped 
using the {\bf -n} option, i.e.
 {\it ./launch  -p param.input -t template.m -n}.  In this case the program will not assume that such a constraint is present, 
 but immediately proceed to the iterated sector decomposition.
 This option can be very useful for cases where the user already has done some mo\-di\-fications to the integrand, 
 for example variable transformations which are convenient for a specific integral at hand, or where one 
 or several Feynman parameters already have been integrated out analytically, or for other non-standard 
 integrals.
 The implementation of this option is very recent and extends the applicability of the program 
 to a much wider class of integrals.
\item The possibility to loop over ranges of numerical values for the Mandelstam invariants, masses ({\tt loop} case) 
or user-defined parameters ({\tt general} case) is automated.
\item To evaluate  parametric functions in the subdirectory {\tt general}, 
the user can define additional (finite) functions at a symbolic level and specify them only later,  
after the integrand has been transformed into a set of finite coefficient integrals for each order in 
$\epsilon$. 
\end{enumerate}

For examples and results, we refer the reader to \cite{Borowka:2012yc} and the {\tt demos} directories 
coming with the program.

\section{Conclusions}

We have presented  \secdec{} version 2, 
an automated program which can be applied to multi-loop integrals
and more general parameter integrals to perform two tasks: 
factorise dimensionally regulated singularities as poles in $1/\eps$  
and numerically 
calculate the coefficients of the resulting Laurent series in $\eps$. 
The program is publicly available at {\it http://secdec.hepforge.org}.

An important new feature of the program is the fact that it now can deal with fully physical 
kinematics, i.e. is not restricted to one-scale problems or Euclidean kinematics anymore.
A new construction of the integrand,  based entirely on topological rules, 
is also included, along with other very useful new features which extend the range of applicability
of the program. 

To calculate full 
two-loop amplitudes involving several mass scales, the timings still leave room for improvement. 
However, the program offers the possibility of major parallelisation if several processors are available. 
An interface to programs performing the reduction to master integrals (not necessarily scalar
integrals), 
which then are fed directly into  \secdec\,, is under construction.


\begin{thebibliography}{10}
\expandafter\ifx\csname url\endcsname\relax
  \def\url#1{{\tt #1}}\fi
\expandafter\ifx\csname urlprefix\endcsname\relax\def\urlprefix{URL }\fi
\providecommand{\eprint}[2][]{\url{#2}}

\bibitem{Carter:2010hi}
Carter J and Heinrich G 2011 {\em Comput.Phys.Commun.\/} {\bf 182} 1566--1581
  (\textit{Preprint} \eprint{1011.5493})

\bibitem{Borowka:2012yc}
Borowka S, Carter J and Heinrich G 2012 {\em Comput. Phys. Commun.\/}
  (\textit{Preprint} \eprint{1204.4152})

\bibitem{Binoth:2000ps}
Binoth T and Heinrich G 2000 {\em Nucl. Phys.\/} {\bf B585} 741--759
  (\textit{Preprint} \eprint{hep-ph/0004013})

\bibitem{Roth:1996pd}
Roth M and Denner A 1996 {\em Nucl. Phys.\/} {\bf B479} 495--514
  (\textit{Preprint} \eprint{hep-ph/9605420})

\bibitem{Hepp:1966eg}
Hepp K 1966 {\em Commun. Math. Phys.\/} {\bf 2} 301--326

\bibitem{Bogner:2007cr}
Bogner C and Weinzierl S 2008 {\em Comput. Phys. Commun.\/} {\bf 178} 596--610
  (\textit{Preprint} \eprint{0709.4092})

\bibitem{Smirnov:2008py}
Smirnov A and Tentyukov M 2009 {\em Comput.Phys.Commun.\/} {\bf 180} 735--746
  (\textit{Preprint} \eprint{0807.4129})

\bibitem{Smirnov:2009pb}
Smirnov A, Smirnov V and Tentyukov M 2011 {\em Comput.Phys.Commun.\/} {\bf 182}
  790--803 (\textit{Preprint} \eprint{0912.0158})

\bibitem{Gluza:2010rn}
Gluza J, Kajda K, Riemann T and Yundin V 2011 {\em Eur.Phys.J.\/} {\bf C71}
  1516 (\textit{Preprint} \eprint{1010.1667})

\bibitem{Soper:1999xk}
Soper D~E 2000 {\em Phys. Rev.\/} {\bf D62} 014009 (\textit{Preprint}
  \eprint{hep-ph/9910292})

\bibitem{Binoth:2002xh}
Binoth T, Heinrich G and Kauer N 2003 {\em Nucl. Phys.\/} {\bf B654} 277--300
  (\textit{Preprint} \eprint{hep-ph/0210023})

\bibitem{Binoth:2005ff}
Binoth T, Guillet J~P, Heinrich G, Pilon E and Schubert C 2005 {\em JHEP\/}
  {\bf 10} 015 (\textit{Preprint} \eprint{hep-ph/0504267})

\bibitem{Nagy:2006xy}
Nagy Z and Soper D~E 2006 {\em Phys. Rev.\/} {\bf D74} 093006
  (\textit{Preprint} \eprint{hep-ph/0610028})

\bibitem{Gong:2008ww}
Gong W, Nagy Z and Soper D~E 2009 {\em Phys. Rev.\/} {\bf D79} 033005
  (\textit{Preprint} \eprint{0812.3686})

\bibitem{Lazopoulos:2007ix}
Lazopoulos A, Melnikov K and Petriello F 2007 {\em Phys. Rev.\/} {\bf D76}
  014001 (\textit{Preprint} \eprint{hep-ph/0703273})

\bibitem{Becker:2010ng}
Becker S, Reuschle C and Weinzierl S 2010 {\em JHEP\/} {\bf 1012} 013
  (\textit{Preprint} \eprint{1010.4187})

\bibitem{Becker:2011vg}
Becker S, Goetz D, Reuschle C, Schwan C and Weinzierl S 2012 {\em
  Phys.Rev.Lett.\/} {\bf 108} 032005 (\textit{Preprint} \eprint{1111.1733})

\bibitem{Kurihara:2005ja}
Kurihara Y and Kaneko T 2006 {\em Comput.Phys.Commun.\/} {\bf 174} 530--539
  (\textit{Preprint} \eprint{hep-ph/0503003})

\bibitem{Anastasiou:2007qb}
Anastasiou C, Beerli S and Daleo A 2007 {\em JHEP\/} {\bf 05} 071
  (\textit{Preprint} \eprint{hep-ph/0703282})

\bibitem{Anastasiou:2008rm}
Anastasiou C, Beerli S and Daleo A 2008 {\em Phys. Rev. Lett.\/} {\bf 100}
  241806 (\textit{Preprint} \eprint{0803.3065})

\bibitem{Freitas:2012iu}
Freitas A 2012 {\em JHEP\/} {\bf 1207} 132 (\textit{Preprint}
  \eprint{1205.3515})

\bibitem{Bauberger:1994by}
Bauberger S, Berends F~A, Bohm M and Buza M 1995 {\em Nucl.Phys.\/} {\bf B434}
  383--407 (\textit{Preprint} \eprint{hep-ph/9409388})

\bibitem{Caffo:2002ch}
Caffo M, Czyz H and Remiddi E 2002 {\em Nucl.Phys.\/} {\bf B634} 309--325
  (\textit{Preprint} \eprint{hep-ph/0203256})

\bibitem{Pozzorini:2005ff}
Pozzorini S and Remiddi E 2006 {\em Comput.Phys.Commun.\/} {\bf 175} 381--387
  (\textit{Preprint} \eprint{hep-ph/0505041})

\bibitem{Czakon:2005rk}
Czakon M 2006 {\em Comput.Phys.Commun.\/} {\bf 175} 559--571 (\textit{Preprint}
  \eprint{hep-ph/0511200})

\bibitem{Czakon:2008zk}
Czakon M 2008 {\em Phys.Lett.\/} {\bf B664} 307--314 (\textit{Preprint}
  \eprint{0803.1400})

\bibitem{Freitas:2010nx}
Freitas A and Huang Y~C 2010 {\em JHEP\/} {\bf 1004} 074 (\textit{Preprint}
  \eprint{1001.3243})

\bibitem{Yuasa:2011ff}
Yuasa F, de~Doncker E, Hamaguchi N, Ishikawa T, Kato K {\em et~al.\/} 2012 {\em
  Comput.Phys.Commun.\/} {\bf 183} 2136--2144 (\textit{Preprint}
  \eprint{1112.0637})

\bibitem{Smirnov:2006ry}
Smirnov V~A 2006 {\em {Feynman integral calculus}\/} (Springer)

\bibitem{Heinrich:2008si}
Heinrich G 2008 {\em Int. J. Mod. Phys.\/} {\bf A23} 1457--1486
  (\textit{Preprint} \eprint{0803.4177})

\bibitem{Hahn:2004fe}
Hahn T 2005 {\em Comput. Phys. Commun.\/} {\bf 168} 78--95 (\textit{Preprint}
  \eprint{hep-ph/0404043})

\bibitem{Agrawal:2011tm}
Agrawal S, Hahn T and Mirabella E 2011  (\textit{Preprint} \eprint{1112.0124})

\bibitem{Kawabata:1995th}
Kawabata S 1995 {\em Comp. Phys. Commun.\/} {\bf 88} 309--326.

\end{thebibliography}

\providecommand{\newblock}{}

\end{document}